\documentclass[a4paper,twoside]{article}

\usepackage{epsfig}
\usepackage{subcaption}
\usepackage{calc}
\usepackage{amssymb}
\usepackage{amstext}
\usepackage{amsmath}
\usepackage{amsthm}
\usepackage{multicol}
\usepackage{pslatex}
\usepackage{apalike}
\usepackage{algorithm2e}
\usepackage[bottom]{footmisc}
\usepackage{SCITEPRESS}     
\usepackage{dblfloatfix}
\usepackage{multirow}
\usepackage{xcolor}

\begin{document}

\title{A Cost-Benefit Analysis of Additive Manufacturing as a Service}

\author{
	\authorname{Igor Ivki\'c\sup{2}\orcidAuthor{0000-0003-3037-7813},
		Tobias Buhmann\sup{3,4} and
		Burkhard List\sup{1}}
	\affiliation{\sup{1}b\&mi, Wiesmath, AT}
	\affiliation{\sup{2}Lancaster University, Lancaster, UK}
	\affiliation{\sup{3}University of Applied Sciences Burgenland / \sup{4}Forschung Burgenland, Eisenstadt, AT}
	\email{i.ivkic@lancaster.ac.uk, tobias.buhmann@outlook.at, b@3d60.works}
	\vspace{-34pt}}

\keywords{Cloud Crafting Platform, Additive Manufacturing, Manufacturing as a Service, Cost-Benefit Analysis}

\abstract{
	The global manufacturing landscape is undergoing a fundamental shift from resource-intensive mass production to sustainable, localised manufacturing. This paper presents a comprehensive analysis of a Cloud Crafting Platform that enables Manufacturing as a Service (MaaS) through additive manufacturing technologies. The platform connects web shops with local three-dimensional (3D) printing facilities, allowing customers to purchase products that are manufactured on-demand in their vicinity. We present the platform's Service-Oriented Architecture (SOA), deployment on the Microsoft Azure cloud, and integration with three different 3D printer models in a testbed environment. A detailed cost-benefit analysis demonstrates the economic viability of the approach, which generates significant profit margins. The platform implements a weighted profit-sharing model that fairly compensates all stakeholders based on their investment and operational responsibilities. Our results show that on-demand, localised manufacturing through MaaS is not only technically feasible but also economically viable, while reducing environmental impact through shortened supply chains and elimination of inventory waste. The platform's extensible architecture allows for future integration of additional manufacturing technologies beyond 3D printing.
}

\onecolumn \maketitle \normalsize \setcounter{footnote}{0} \vfill

\section{\uppercase{Introduction}}
\label{sec:sec1}

The global manufacturing landscape has long been characterised by resource-intensive production processes, extended supply chains, and a tendency towards overproduction \cite{ref001}. These practices not only contribute to significant environmental impacts, but also create economic dependencies that can be disrupted by global events, as demonstrated by the COVID-19 pandemic \cite{ref002} or the Suez Canal blockade \cite{ref003,ref004}. In addition, there is a growing consumer preference for fair, local, and sustainable products \cite{ref005}, indicating a shift in market demands that traditional production models struggle to meet.

The integration of cloud computing and additive manufacturing technologies has opened up new opportunities to transform traditional manufacturing processes. This convergence has given rise to the concept of Manufacturing as a Service (MaaS), which has the potential to transform production methods, supply chains, and customer experiences in the long term. Our research is focused on developing a cloud-based platform that connects webshop owners with small and medium-sized enterprises (SMEs) that operate three-dimensional (3D) printers, enabling on-demand, localised production of goods \cite{ref006}.

In this paper, we propose a Cloud Crafting Platform that addresses the challenges of mass production and the traditional supply chain by enabling a paradigm shift in how products are manufactured and distributed. The proposed platform acts as a bridge between customers, online retailers, and local 3D printing facilities, enabling the production of goods only after they have been purchased. This on-demand approach not only reduces waste and inventory costs, but also empowers customers to actively participate in the manufacturing process by choosing local, sustainable production methods.

The platform's architecture is designed to be scalable and flexible, using a serverless approach that can seamlessly connect web shops (\textbf{\textit{points of sale}}) with 3D printer operators (\textbf{\textit{points of manufacture}}). This approach not only stimulates local economic growth, but also significantly reduces the environmental impact associated with long-distance shipping and traditional mass production techniques.

Our research goes beyond the theoretical framework to include a comprehensive cost-benefit analysis that evaluates the economic viability of this MaaS approach. By comparing different 3D printer models and analysing metrics such as print time, material usage, and energy consumption, we provide insights into the operational efficiencies and potential cost savings of on-demand, localised manufacturing.

This paper extends on our previous work \cite{ref007}, which introduced the concept of a MaaS platform for on-demand, localised manufacturing using 3D printing technology. Building on this foundation, this paper presents a comprehensive analysis of our Cloud Crafting Platform, presenting its architectural design, addressing implementation challenges, and providing detailed results of our cost-benefit analysis. This paper provides a deeper exploration of the platform's potential to transform manufacturing processes and traditional supply chains.

By exploring the intersection of cloud computing \cite{ref008}, additive manufacturing \cite{ref009}, and on-demand production \cite{ref006}, our research contributes to the ongoing dialogue about the future of manufacturing. We discuss how MaaS can not only improve production efficiency and reduce environmental impact, but also support local economic development and increase supply chain resilience. This work has implications for a range of stakeholders, including manufacturers, retailers, consumers, and policymakers, as we move towards a more sustainable and responsive manufacturing ecosystem.

The remainder of this paper is structured as follows: Section \ref{sec:sec2} identifies related work in the field and presents the theoretical foundations and technological advances underlying additive MaaS. Next, in Section \ref{sec:sec3}, the Cloud Crafting Platform prototype design, architecture, and testbed setup are presented. Based on that, Section \ref{sec:sec4} presents the results of the cost-benefit analysis, highlighting key metrics and their implications for the feasibility of the approach including a discussion of the results. Finally, Section \ref{sec:sec5} concludes with a summary of the contributions and directions for future work.

\section{\uppercase{Related Work}}
\label{sec:sec2}

This chapter provides a comprehensive overview of the current state of research in three key areas relevant to our study: (1) cloud-based manufacturing systems, (2) additive manufacturing, and (3) the associated business models, sustainability, and social aspects. By examining these interrelated areas, we aim to provide a solid foundation for our research and identify the gaps that our work will address.

\subsection{Cloud-based Manufacturing}

Cloud-based manufacturing systems represent a significant advancement in the manufacturing landscape, often using a combination of advanced technologies such as cloud computing, the Internet of Things (IoT) and artificial intelligence (AI) to improve the flexibility and efficiency of traditional manufacturing processes. Thames and Schaefer (2017)\nocite{ref010} provide an overview of the technological foundations for cloud manufacturing, discussing both the challenges and opportunities presented by these advanced technologies. Caiazzo et al. (2022)\nocite{ref011} demonstrate the potential for improved process monitoring and control through AI-assisted monitoring and risk classification in manufacturing environments.

Zhang et al. (2014)\nocite{ref012} explore the feasibility of cloud manufacturing architectures, presenting a practical prototype that provides a comprehensive and integrated platform for the manufacturing industry. Building on this, Cui et al. (2022)\nocite{ref013} propose a model for 3D printing in cloud manufacturing, outlining four different roles: cloud operators, 3D printing service providers, demanders, and logistics service providers.

Škulj et al. (2017)\nocite{ref014} discuss a distributed network architecture that provides a more flexible and scalable alternative to centralised systems, supported by compute and knowledge clouds. Lu et al. (2014)\nocite{ref015} explore the concept of a hybrid manufacturing cloud, which combines traditional manufacturing processes with cloud-based technologies to enable efficient integration and resource optimisation.

Wang et al. (2019)\nocite{ref016} and Rudolph and Emmelmann (2017)\nocite{ref017} explore the integration and optimisation of manufacturing resources through cloud computing, enabling the creation of interoperable cloud manufacturing concepts that support the global optimisation of manufacturing resources. They highlight how cloud manufacturing can enhance collaborative manufacturing environments through virtualisation and services.

Adamson et al. (2017)\nocite{ref018} and Wu et al. (2013)\nocite{ref019} address the challenges associated with cloud manufacturing, including cybersecurity concerns and the need for intelligent monitoring and control systems. They also discuss future developments and potential solutions to fully realise the full potential of cloud manufacturing.

Giunta et al. (2023)\nocite{ref020}, Vedeshin et al. (2020)\nocite{ref021} and Simeone et al. (2020)\nocite{ref022} present innovative applications and forward-looking technologies in cloud manufacturing. Their work illustrates how cloud-based technologies are transforming the manufacturing industry by offering flexibility and personalised manufacturing capabilities.

\subsection{Additive Manufacturing Systems}

Additive manufacturing (or commonly known as 3D printing) has seen significant technological advances in recent years and has the potential to transform the way how products are made across multiple industries by challenging and complementing traditional manufacturing approaches. 

Shahrubudin et al. (2019)\nocite{ref023} and Rauch et al. (2018)\nocite{ref024} provide a comprehensive overview of the transformative impact of additive manufacturing on industrial applications such as automotive, aerospace, and mechanical engineering. They highlight the ability of 3D printing techniques to enable the production of complex parts that are difficult or impossible to manufacture using traditional methods. The authors also discuss the benefits in terms of reducing material waste and accelerating design cycles, leading to shorter iteration and innovation cycles.

The application of additive manufacturing in military contexts demonstrates the use of this technology to overcome logistical challenges and increase efficiency. Jagoda et al. (2020)\nocite{ref025} describe the use of 3D printing for rapid on-site production of parts, which is particularly beneficial in conflict and crisis areas. Fiske et al. (2018)\nocite{ref026} discuss the possibility of using additive manufacturing to construct building structures through additive manufacturing in remote areas. Rankin et al. (2014)\nocite{ref027} highlight the cost-effectiveness and functionality of 3D printed surgical instruments produced through 3D printing, which can improve medical care on the battlefield.

In the medical field, Url et al. (2022)\nocite{ref028} discuss the production of personalised medical implants and surgical instruments through 3D printing. They highlight how 3D printing services are being integrated into hospitals, leading to improved, patient-specific treatment strategies. Ghilan et al. (2020)\nocite{ref029} provide a comprehensive overview of developments in 3D and 4D printing, highlighting the importance of machine learning in improving design efficiency and optimising the functionality of medical devices.

Panda et al. (2023)\nocite{ref030} discuss technical limitations such as quality assurance in additive manufacturing processes and challenges in material selection. They emphasise that scaling up production and integration into existing manufacturing systems must be strategically planned to achieve effective results.

\subsection{Business Models, Sustainability and Social Aspects}

The rapid development of digital technologies and their integration into the manufacturing industry is leading to fundamental changes in traditional business models. The additive MaaS approach exemplifies this shift, allowing companies to control and scale their production processes via cloud-based platforms. Goldhar and Jelinek (1990)\nocite{ref031} emphasise that Computer Integrated Manufacturing (CIM) improves product variety and customisation by promoting the integration of information technology into production processes. This more flexible production environment allows companies to respond quickly and cost-effectively to individual customer requirements.

Nie et al. (2023)\nocite{ref032} discuss the flexibility of 3D printing in monopolistic markets and its impact on customer loyalty and market dominance of companies. Ivanov et al. (2022)\nocite{ref033} further explain that cloud supply chain models enable seamless integration and management of physical and digital assets, increasing operational flexibility and more dynamic adaptation of resources to changing market conditions.

Rauch et al. (2018)\nocite{ref024} and Smith et al. (2013)\nocite{ref034} discuss the impact of the introduction of mass customisation and adaptive manufacturing systems on traditional production paradigms. The introduction of these technologies implies not only a change in manufacturing processes, but also a redesign of customer interactions and the entire value chain.

Pahwa and Starly (2021)\nocite{ref035} show how the use of deep reinforcement learning can improve decision making on MaaS platforms, increasing adaptability and responsiveness. Sun et al. (2024)\nocite{ref036} provide a detailed analysis showing that, under certain conditions, additive manufacturing processes can outperform traditional methods in terms of cost and speed, particularly for rapid product iterations and complex designs.

Chaudhuri et al. (2021)\nocite{ref037} evaluate the impact of differentiated pricing strategies on MaaS platforms on profitability and market penetration. Strategic pricing decisions allow companies to tailor their offerings to specific market segments and thereby achieve optimal market positioning. 

Dhir et al. (2023)\nocite{ref038} and Bulut et al. (2021)\nocite{ref039} examine the challenges and potential of MaaS for SMEs in different geographical regions. Tao et al. (2017)\nocite{ref040} and Fisher et al. (2018)\nocite{ref041} discuss the environmental aspects of sustainability in manufacturing. They show how cloud manufacturing can improve resource efficiency, minimise waste, and increase energy efficiency through optimised manufacturing processes and improved material usage.

\subsection{Summary}
\begin{table*}[b]
	\caption{Summary of Related Work on Additive MaaS.}
	\resizebox{\textwidth}{!}{%
		\begin{tabular}{|l|l|}
			\hline
			\textbf{Cloud-based Manufacturing Systems}                                                       & \multicolumn{1}{c|}{\textbf{Sources}} \\ \hline
			\multirow{2}{*}{Technological foundations and advanced technologies}                             & \textit{Thames and Schaefer (2017),}  \\ 
			& \textit{Caiazzo et al. (2022)}        \\ \hline
			\multirow{2}{*}{Feasibility and integrated platforms in the context of Cloud Manufacturing}      & \textit{Zhang et al. (2014),}         \\ 
			& \textit{Cui et al. (2022)}            \\ \hline
			\multirow{2}{*}{Decentralised and hybrid cloud network architectures to increase flexibility}    & \textit{Škulj et al. (2017),}         \\ 
			& \textit{Lu et al. (2014)}             \\ \hline
			\multirow{2}{*}{Interoperability and service-oriented Cloud Manufacturing systems}               & \textit{Wang et al. (2019),}          \\ 
			& \textit{Rudolph and Emmelmann (2017)} \\ \hline
			\multirow{2}{*}{Cybersecurity challenges and development of monitoring systems}                  & \textit{Adamson et al. (2017),}       \\ 
			& \textit{Wu et al. (2013)}             \\ \hline
			\multirow{3}{*}{Innovative applications and future-oriented technologies}                        & \textit{Giunta et al. (2023),}        \\ 
			& \textit{Vedeshin et al. (2020),}      \\ 
			& \textit{Simeone et al. (2020)}        \\ \hline
			\textbf{Additive Manufacturing}                                                                  & \multicolumn{1}{c|}{\textbf{Sources}} \\ \hline
			\multirow{2}{*}{Transformative impact on industrial applications through additive manufacturing} & \textit{Shahrubudin et al. (2019),}   \\ 
			& \textit{Rauch et al. (2018)}          \\ \hline
			\multirow{3}{*}{Additive manufacturing and its benefits in military applications}                & \textit{Jagoda et al. (2020),}        \\ 
			& \textit{Fiske et al. (2018),}         \\ 
			& \textit{Rankin et al. (2014)}         \\ \hline
			\multirow{2}{*}{Additive manufacturing and its benefits in medical applications}                 & \textit{Url et al. (2022),}           \\ 
			& \textit{Ghilan et al. (2020)}         \\ \hline
			\multirow{3}{*}{Challenges in implementing additive manufacturing}                               & \textit{Panda et al. (2023),}   \\ 
			& \textit{Url et al. (2022),}           \\ 
			& \textit{Ghilan et al. (2020)}         \\ \hline
			\textbf{Business Models, Sustainability and Social Aspects}                                      & \multicolumn{1}{c|}{\textbf{Sources}} \\ \hline
			\multirow{2}{*}{Business model innovations and their impact on market structures}                & \textit{Goldhar and Jelinek (1990),}  \\ 
			& \textit{Nie et al. (2023)}            \\ \hline
			\multirow{3}{*}{Cloud supply chain models and mass customisation}                                & \textit{Ivanov et al. (2022),}        \\ 
			& \textit{Rauch et al. (2018),}         \\ 
			& \textit{Smith et al. (2013)}          \\ \hline
			\multirow{3}{*}{Optimisation of MaaS platforms and economic analyses}                            & \textit{Pahwa and Starly (2021),}     \\ 
			& \textit{Sun et al. (2024),}           \\ 
			& \textit{Chaudhuri et al. (2021)}      \\ \hline
			\multirow{4}{*}{Regional adaptations and sustainability approaches for Cloud Manufacturing}      & \textit{Dhir et al. (2023),}          \\ 
			& \textit{Bulut et al. (2021),}         \\ 
			& \textit{Tao et al. (2017),}           \\ 
			& \textit{Fisher et al. (2018)}         \\ \hline
		\end{tabular}%
	}
	\label{tab:tab1}
\end{table*} 

The identified literature reveals a growing body of research on cloud-based manufacturing systems, additive manufacturing, and their associated business models and sustainability aspects (as shown in Table \ref{tab:tab1}). However, there are still significant gaps in our understanding of how these technologies can be effectively integrated and implemented in real-world scenarios, particularly in the context of MaaS. 

While existing research explores the benefits and challenges of cloud-based manufacturing systems and additive manufacturing, few studies focus on the integration of these technologies within the MaaS ecosystem and their practical applicability in different industrial contexts. In particular and to the best of our knowledge, the area of designing and implementing a prototype that provides additive MaaS has not been explored in detail. 

This paper aims to address these gaps by developing and analysing a MaaS-based prototype that integrates cloud-based systems with additive manufacturing. Our work contributes to the field by (1) providing a comprehensive analysis of the architectural and technical requirements for implementing additive manufacturing as a cloud-based service, (2) performing a direct comparison of cost structures between different 3D printers within a MaaS approach, (3) evaluating the impact of an additive MaaS approach on reducing environmental impact and promoting regional economic development in the context of sustainable production practices, and (4) developing and analysing a functional prototype to demonstrate the practical implementation of these concepts.
\begin{figure*}[!b] 
	\centering
	\includegraphics[width=\textwidth]{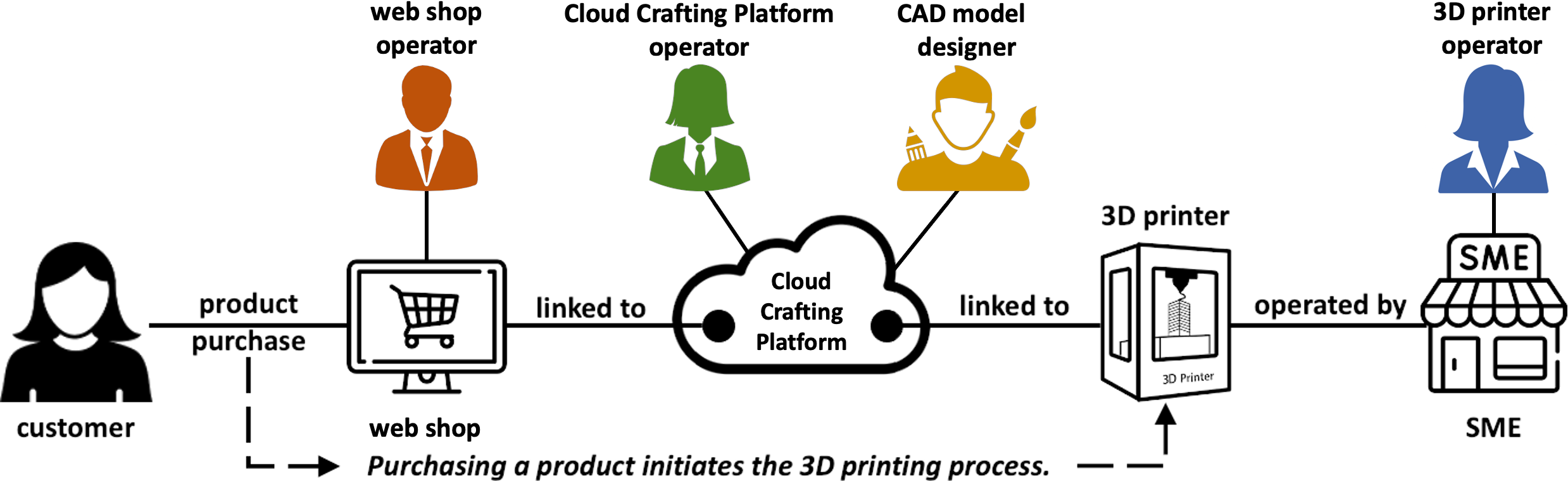}
	\caption{Overall Use Case where a buying customer initiates the on-demand MaaS process (adapted from Ivki\'c et al., 2024).} 
	\label{fig:fig1}
\end{figure*}

\section{\uppercase{Cloud Crafting Platform}}
\label{sec:sec3}

\begin{figure*}[!b] 
	\centering
	\includegraphics[width=\textwidth]{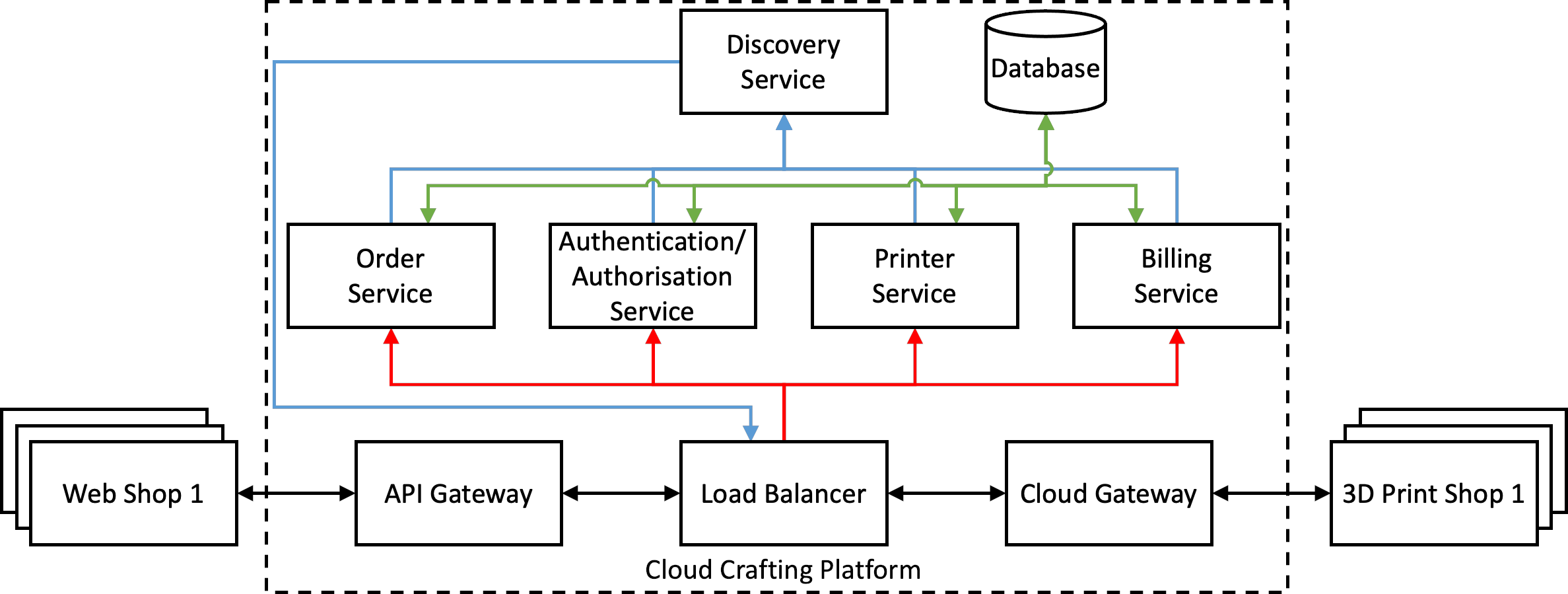}
	\caption{Service-Oriented Cloud Crafting Platform Architecture (adapted from Ivki\'c et al., 2024).} 
	\label{fig:fig2}
\end{figure*}

Customer attitudes have changed to the point where they want to know more than ever where their products come from and how they are made. The desire to actively influence the manufacturing process is greater than ever. Many prefer to support local economies and buy regionally and sustainably produced products rather than rely on imported mass-produced goods. Thanks to technological advances in additive manufacturing \cite{ref009}, it is now possible to produce a wide range of products \cite{ref023} locally and cost-effectively using three-dimensional (3D) printing technologies \cite{ref042}. At the heart of the MaaS approach is a Cloud Crafting Platform that can be used by 3D printer operators to integrate their 3D printers online and offer 3D printing as a service. 

With this Cloud Crafting Platform, any 3D printer operator can become a local on-demand producer, capable of handling a certain amount of customisation during 3D printing (or manufacturing). This strengthens the local economy and offers customers the opportunity to buy products that have been made in the nearby region. Another aim of the Cloud Crafting Platform is to break up traditional supply chains and produce products where they are bought. This eliminates long transport routes and has a positive impact on the environment. 

The Cloud Crafting Platform follows a serverless architecture approach that can connect both web shops and 3D printer operators. the following sections outline the overall use case, prototype architecture, and experimental testbed setup of the Cloud Crafting Platform, which aims to provide an alternative to traditional manufacturing processes by enabling local, sustainable production.

\subsection{Overall Use Case}

As shown in Figure \ref{fig:fig1}, the Cloud Crafting Platform orchestrates a comprehensive ecosystem connecting five key stakeholders in an innovative additive MaaS approach. At its core, the platform acts as a bridge between the \textbf{\textit{point of sale}} (web shops) and the \textbf{\textit{point of manufacture}} (3D printer operators), enabling a seamless on-demand production process. Similar to how mobile app stores connect developers and users, the Cloud Crafting Platform maintains a repository of Computer Aided Design (CAD) models created by 3D model designers. These digital blueprints form the basis of the platform's product catalogue. Before the products can be offered in the web shop, the CAD model designers upload their 3D designs to the platform, where they become available for commercial use. The web shop operators can then integrate these designs into their online stores, offering products that do not yet exist in physical form, but can be manufactured on demand.

When a customer makes a purchase, it triggers an order workflow that initiates the manufacturing process through the web shop, which forwards the request to the Cloud Crafting Platform. The platform then identifies the nearest available SME with appropriate 3D printing capabilities and transmits the production specifications. This SME, acting as the 3D printer operator, manufactures the product locally, significantly reducing transport distances and supporting regional economic development. The architecture of the Cloud Craftin Platform enables five different stakeholders to interact seamlessly within the additive MaaS ecosystem, namely (\textbf{Stakeholder 1}) the customer, who initiates the process by purchasing a product, (\textbf{Stakeholder 2}) the web shop operator who provides the \textbf{\textit{point of sale}}, (\textbf{Stakeholder 3}) the Cloud Crafting Platform operator who manages the overall service and enables the interaction of all stakeholders involved, (\textbf{Stakeholder 4}) the CAD model designer who creates the digital product designs, and (\textbf{Stakeholder 5}) the 3D printer operator (SME) who produces the physical products.

The Cloud Crafting Platform goes beyond simply connecting the identified stakeholders; it also creates a profit-sharing ecosystem where revenue from each product sold is shared among the four key service providers (web shop operator, Cloud Crafting Platform operator, CAD model designer, 3D printer operator). This ensures that each stakeholder receives fair compensation for their contribution to the manufacturing process \cite{ref007}. 

Furthermore, while the initial implementation of the Cloud Crafting Platform focuses on 3D printing technology, its architecture is designed to accommodate a wider range of manufacturing technologies. The system can be extended to integrate various manufacturing machines such as Computerised Numerical Control (CNC), laser cutting, plotter cutting, robotics, and augmented reality technologies
, making it a versatile solution for different manufacturing needs. This extensibility ensures that the platform can evolve with technological advances and adapt to changing market demands while maintaining its core principle of connecting customers with local manufacturers.

\begin{figure*}[!b] 
	\centering
	\includegraphics[width=\textwidth]{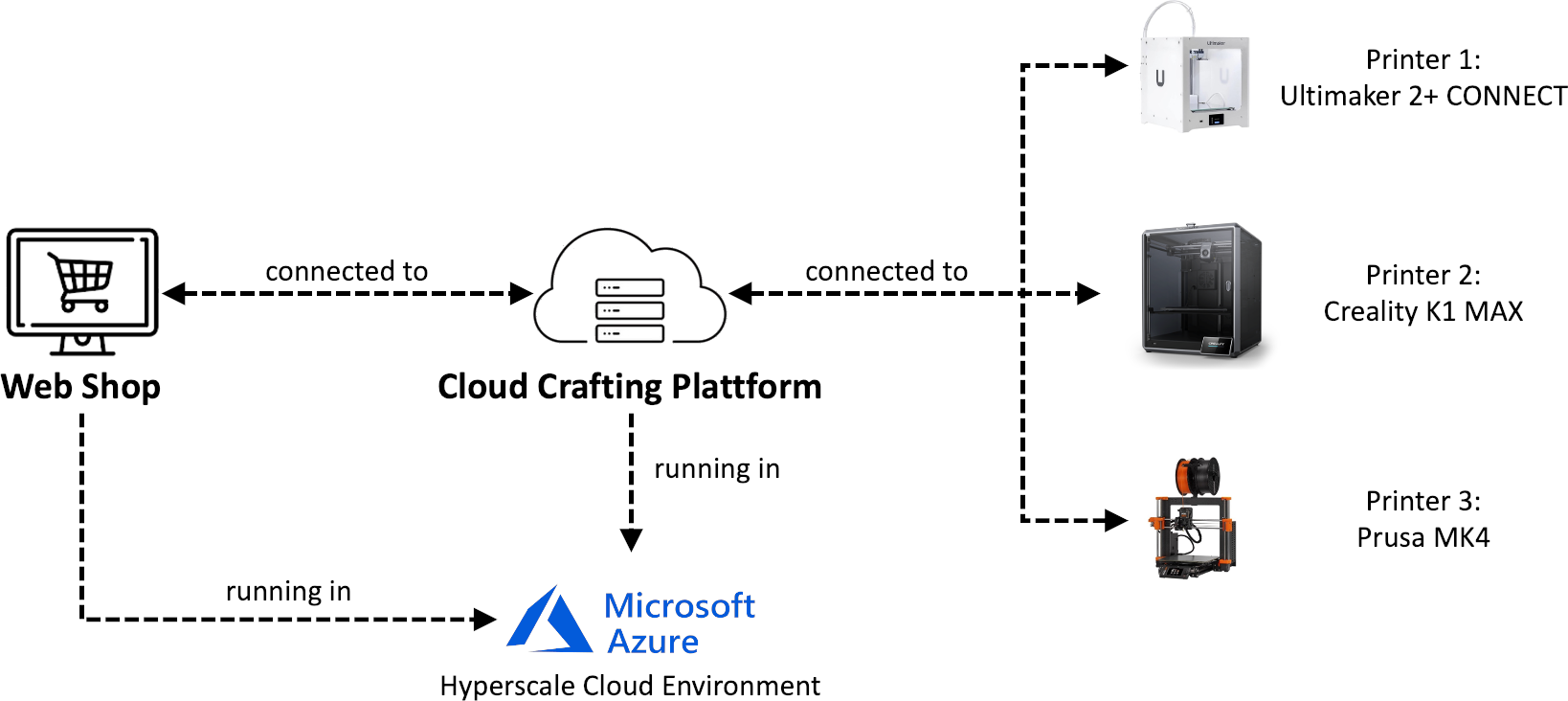}
	\caption{Testbed Setup for End-to-End Validation of the MaaS Approach.} 
	\label{fig:fig3}
\end{figure*}
\subsection{Prototype Architecture}

The Cloud Crafting Platform follows a Service-Oriented Architecture (SOA) designed to ensure scalability, reliability, and seamless integration between web shops and 3D print shops. The architecture consists of two gateways, a central load balancer, and five core services that work together to enable the MaaS ecosystem. Together, all the building blocks provide the necessary endpoints to connect all the web shops and 3D print shops, including all the services, to process a purchased product and route it to the nearest available MaaS production site. To achieve this goal, the platform provides two gateways for external shop and production site integration and external communication:
\begin{itemize}
	\item\textbf{API Gateway}: this gateway acts as an entry point for web shops, handling all incoming requests and ensuring secure communication protocols.
	\item\textbf{Cloud Gateway}: this gateway manages connections with print shops and enables real-time communication with the production site equipment.
\end{itemize}

The \textbf{\textit{Load Balancer}} acts as the primary traffic manager, distributing incoming requests across services to ensure optimal resource utilisation and high availability. By working in conjunction with the \textbf{\textit{Discovery Service}}, which enables dynamic service registration and discovery, these two services eliminate the need for hard-coded service locations. 

The \textbf{\textit{Order Service}} manages the complete lifecycle of manufacturing orders from inception to completion. It handles the creation of new orders as customers make purchases, maintains real-time order status updates, and stores all order-related information. This service ensures that each order is properly tracked and executed throughout the manufacturing process.

The \textbf{\textit{Authentication/Authorisation Service}} is the security backbone of the platform, managing user registration processes, handling login requests, and validating security tokens for all platform interactions. This service ensures that only authorised users have access to specific platform features and maintains secure communication channels between different components of the system. It implements role-based access control to ensure that each stakeholder only has access to the functionality relevant to their role.

The \textbf{\textit{Printer Service}} orchestrates all aspects of the manufacturing process, managing the scheduling and execution of print jobs, monitoring the real-time status of connected 3D printers, and maintaining printer availability information. This service also manages the queuing of print jobs, ensuring optimal use of available printing resources while maintaining quality control throughout the manufacturing process.

The \textbf{\textit{Billing Service}} handles all financial aspects of the platform, processing transactions, managing the platform's profit-sharing mechanism between stakeholders, and handling the creation, validation, and redemption of promotional codes. This service maintains detailed records of all financial transactions and ensures the distribution of revenue between the web shop operator, Cloud Crafting Platform operator, CAD model designer, and 3D printer operator.

A centralised database system underpins these services, storing essential information including (1) user profiles and authentication data, (2) order details and status, (3) printer configurations and availability, (4) billing and transaction records, and (5) redeem codes and usage history. All in all, this architecture ensures that the Cloud Crafting Platform can efficiently handle the complex interactions between customers, web shops, and MaaS production sites while maintaining security, scalability, and reliability.

\subsection{Testbed Setup}

To perform a cost-benefit analysis, the overall use case from Figure \ref{fig:fig1} was implemented following the prototype architecture described in the previous section. This allows an end-to-end validation of the Cloud Crafting Platform functionality from product purchase to local manufacturing. In addition to that the testbed can be used to evaluate the production costs associated with manufacturing a given product using the MaaS approach.

As shown in Figure \ref{fig:fig3}, the Cloud Crafting Platform was deployed in the Microsoft Azure cloud, implementing all of the core services and components described in the prototype architecture. This cloud deployment hosts the \textbf{\textit{Load Balancer}}, the \textbf{\textit{Discovery Service}}, and the four core services (\textbf{\textit{Order, Authentication/Authorization, Printer, and Billing}}), as well as the central database. In addition, a web shop was implemented and deployed on a dedicated virtual machine to serve as the \textbf{\textit{point of sale}}. This web shop was integrated with the Cloud Crafting Platform through the \textbf{\textit{API Gateway}}, allowing for seamless transfer of purchase orders to the cloud environment for processing by the platform's core services.

The \textbf{\textit{point of manufacture}} was set up in a laboratory environment, simulating a local SME production site, including the following three networked 3D printers:
\begin{itemize}
	\item\textbf{Printer 1}: Ultimaker 2+ CONNECT
	\item\textbf{Printer 2}: Creality K1 MAX
	\item\textbf{Printer 3}: Prusa MK4\\
\end{itemize}

Each 3D printer was connected to a dedicated Raspberry Pi running the OctoPi operating system (OS), which provides (1) remote printer control capabilities, (2) real-time monitoring of print jobs, (3) secure communication with the Cloud Crafting Platform via the \textbf{\textit{Cloud Gateway}}, and (4) integration with the platform's \textbf{\textit{Printer Service}} for job management. The Raspberry Pi devices act as local controllers, managing printer operations and maintaining bi-directional communication with the Cloud Crafting Platform via the \textbf{\textit{Cloud Gateway}}. This setup enables automated order processing, printer control, and status monitoring, completing the end-to-end manufacturing process from online purchase to local production. The following figure shows the local setup of a 3D printer connected to a Raspberry Pi that runs OctoPi OS interacting with the Cloud Crafting Platform:
\vspace{-10pt}
\begin{figure}[h]
	\centering
	\includegraphics[width=\columnwidth]{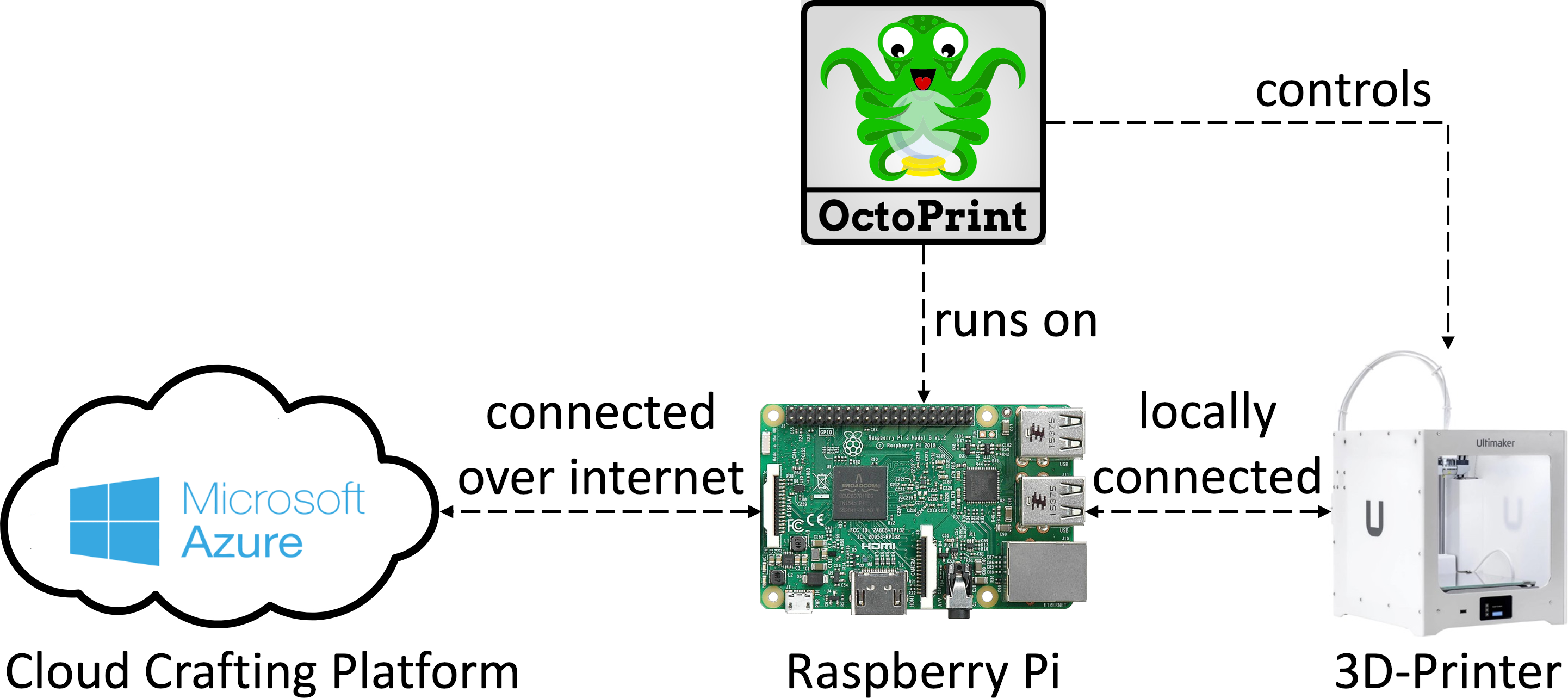}
	\caption{Testbed Laboratory Setup Simulating a Local SME Production Site.} 
	\label{fig:fig4}
\end{figure}

\vspace{-20pt}
\section{\uppercase{Cost-Benefit Analysis}}
\label{sec:sec4}

To evaluate the economic viability of the Cloud Crafting Platform, we conducted a comprehensive cost-benefit analysis using a designed ring as a test product. This analysis evaluates the operational costs across three key dimensions: web shop hosting, cloud platform operation, and actual production costs. Our analysis focused on a specifically designed ring that was uploaded to the deployed Cloud Crafting Platform and made available for purchase through the integrated web shop. The following figure shows a designed test product (ring) for the cost-benefit analysis:
\vspace{-10pt}
\begin{figure}[h]
	\centering
	\includegraphics[width=2.4cm]{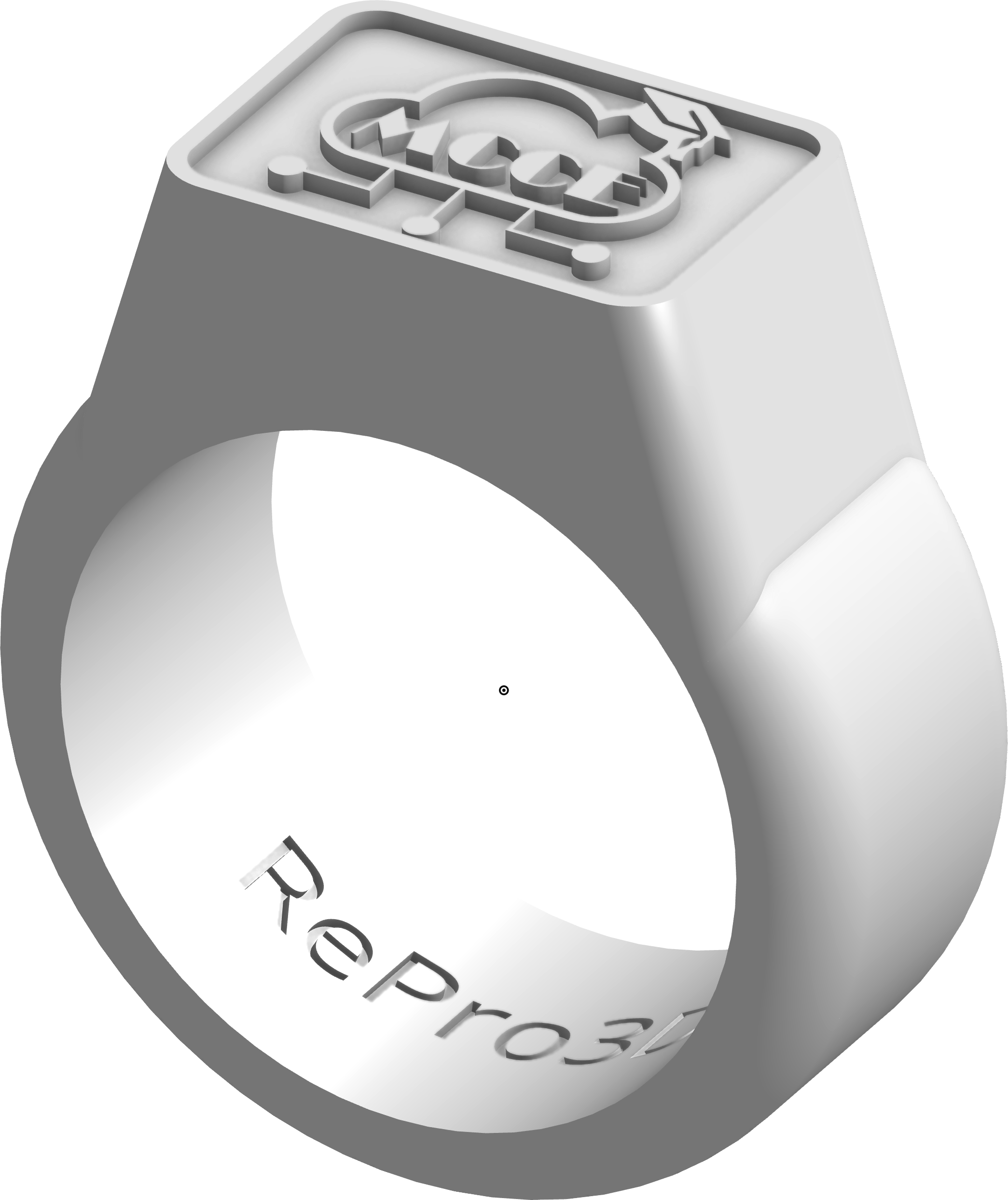}
	\caption{Designed Ring as a Test Product for the Cost-Benefit Analysis.} 
	\label{fig:fig5}
\end{figure}

The cost-benefit analysis included three primary cost categories that are essential for evaluating the economic viability of the Cloud Crafting Platform. The \textbf{\textit{Web Shop Operational Costs}} include all expenses associated with maintaining an online retail presence through Shopify. This includes the monthly subscription fees for hosting the e-commerce platform, transaction processing fees for each sale, and the technical integration costs required to connect the web shop with the Cloud Crafting Platform. Additional costs include domain registration, Secure Sockets Layer (SSL) certificates, and any premium features or plug-ins required to run the online storefront.

The \textbf{\textit{Microsoft Azure Cloud Infrastructure Costs}} are a significant component of the operational expenses. These costs include compute resources for running the microservices architecture, storage capacity for maintaining the CAD model repository, and bandwidth consumption for data transfer between different platform components. The analysis includes Azure service fees across multiple components, including the \textbf{\textit{API Gateway}}, the \textbf{\textit{Load Balancer}}, the \textbf{\textit{Discovery Service}}, and the four core services (\textbf{\textit{Order, Authentication/Authorization, Printer,}} and \textbf{\textit{Billing}}). Additionally, the costs include database hosting, monitoring tools, and security services required to maintain a robust and reliable cloud platform.

The \textbf{\textit{Production Costs}} represent the actual manufacturing expenses incurred at the local SME level. This category includes three critical components: First, the temporal costs measure the complete production cycle time, from initial setup and printer preparation to the final post-processing steps. Second, material consumption costs include both the primary printing material and any support structures required during the manufacturing process. Thirdly, energy consumption costs are evaluated using Shelly Plug devices, which measure the power consumption of both the 3D printers and their controlling Raspberry Pi. These measurements provide a comprehensive view of the resource requirements and associated costs for local, on-demand manufacturing.

To evaluate the economic viability of the Cloud Crafting Platform, we conducted a systematic cost-benefit analysis through repeated test runs of the entire manufacturing process. Using our deployed web shop, we initiated three simultaneous purchase orders for identical rings, which were processed through the Cloud Crafting Platform and distributed to our three different 3D printers (Ultimaker 2+ CONNECT, Creality K1 MAX, and Prusa MK4). This parallel production setup allowed us to directly compare the manufacturing performance and associated costs between different printer models. To ensure statistical significance and to account for potential variations in the manufacturing process, we performed 50 complete test runs, resulting in each printer producing 50 identical rings (n=50). This comprehensive approach allowed us to collect data on production times, material usage, and energy consumption for different printer models while operating under identical conditions.

\subsection{Evaluation}

The material cost to produce a single ring varied between the three 3D printers based on their respective filament prices and usage. The Ultimaker 2+ CONNECT (\textbf{Printer 1}), which uses 2.9 grams (g) of filament at a cost of € 42.99 per 750g, had a material cost of € 0.17 per ring. The Creality K1 MAX (\textbf{Printer 2}), which uses 2.835g of filament at a cost of € 23.14 per kilogram (kg), had a material cost of € 0.07 per ring. Similarly, the Prusa MK4 (\textbf{Printer 3}), which uses 2.85g of filament at a cost of € 29.99 per kg, had a material cost of € 0.09 per ring. 

While all three printers used comparable amounts of filament (between 2.835g and 2.9g), the Creality K1 MAX (\textbf{Printer 2}) proved to be the most cost-effective in terms of material usage, followed by the Prusa MK4 (\textbf{Printer 3}), while the Ultimaker 2+ CONNECT (\textbf{Printer 1}) had the highest material cost per ring, primarily due to its more expensive filament. For each printer, the material cost per ring can be expressed using the following equations:
\begin{equation}\label{eq:eq1}
	C_{material,printer} = \frac{m_{ring}}{m_{spool}} * p_{spool}
\end{equation}
\noindent where:
\begin{table}[h]
	\resizebox{\columnwidth}{!}{%
		\begin{tabular}{ll}
			$m_{ring}$    & is the mass of filament used per ring \\
			$m_{spool}$   & is the mass of filament per spool     \\
			$p_{spool}$   & is the price per spool                \\
			$C_{material, printer}$ & is material cost printer for each ring
		\end{tabular}%
	}
\end{table}

The following equations calculate the material costs for each printer:

$C_{material,printer 1} = \frac{2.9g}{750g} * $ € $42.99 =$  € $0.17$
\vspace{5pt}

$C_{material, printer 2} = \frac{2.835g}{1000g} * $ € $23.14 = $ € $0.07$
\vspace{5pt}

$C_{material,printer 3} = \frac{2.85g}{1000g} * $ € $29.99 = $ € $0.09$
\vspace{10pt}

To provide a detailed insight into energy consumption patterns, the power consumption measurements were divided into three distinct phases. The "Pre-Print" phase includes the initial warm-up period during which the printer heats both the nozzle and the build plate to the required temperatures before printing can begin. The "Print" phase is the actual production period during which the printer actively 3D prints the ring. Finally, the "Post-Print" phase captures the energy consumed during the cool-down period when the printer returns to its standby state at the end of the production process. This three-phase measurement approach provides a comprehensive understanding of energy consumption patterns throughout the entire manufacturing cycle and allows for more accurate cost calculations of the additive manufacturing process. The following equation calculates the power consumption costs for each printer:
\begin{equation}
	C_{E,printer} = \frac{(W_{printer} + W_{pi})}{1000} * T_{E}
\end{equation}

\noindent where:
\begin{table}[h]
	\resizebox{\columnwidth}{!}{%
		\begin{tabular}{ll}
			\multirow{2}{*}{$W_{printer}$}    & represents the power consumption of the \\
			& 3D printer in Watt hours (Wh)\vspace{3pt}\\
			
			\multirow{2}{*}{$W_{pi}$}   & represents the power consumption of the \\
			& Raspberry Pi in Wh\vspace{3pt}\\
			
			$\frac{x}{1000}$   & converts Wh to kWh\vspace{3pt} \\
			
			\multirow{2}{*}{$T_{E}$} & represents the energy rate in Euro (€)/kWh\\ 
			& (we used 0.30 €/kWh for the experiment)\vspace{3pt} \\ 
			
			\multirow{2}{*}{$C_{E,printer}$} & represents the total energy cost in €\\
			& for each printer setup
		\end{tabular}%
	}
\end{table}

Similar to the energy consumption measurements, the printing process time is also divided into the three different operational phases (Pre-Prin, Print, and Post-Print) to enable more accurate cost calculations. These time measurements can be converted into monetary costs by taking into account the power consumption during each phase. The total time-based energy cost can be calculated using the following equation:
\begin{equation}\label{eq:eq3}
	C_{time} = \frac{(t_{pre}*P_{pre} + t_{print}*P_{print} + t_{post}*P_{post})}{1000} * T_{E}
\end{equation}

\noindent where:
\begin{table}[h]
	\resizebox{\columnwidth}{!}{%
		\begin{tabular}{ll}
			\multirow{2}{*}{$t_{pre}, t_{print}, t_{post}$}    & represents the duration  \\
			& of each phase in hours\vspace{3pt}\\
			
			\multirow{3}{*}{$P_{pre}, P_{print}, P_{post}$}   & represent the total power \\
			& consumption in watts (W) \\
			& during each respective phase\vspace{3pt}\\
			
			$\frac{x}{1000}$   & converts Wh to kWh\vspace{3pt} \\
			
			\multirow{3}{*}{$T_{E}$} & represents the energy rate \\ 
			& in Euro (€)/kWh (we used \\
			& 0.30 €/kWh for the experiment)\vspace{3pt} \\ 
		\end{tabular}%
	}
\end{table}

\begin{table*}[!t]
	\caption{Summary of Production Costs per 3D-Printer to manufacture the Test Product (Ring).}\centering
	\resizebox{10cm}{!}{%
		\begin{tabular}{lccccc}
			& Phase                           & Time                          & Power (W)                  & Material (€)              & Total (€)                                   \\ \hline
			\multicolumn{1}{|l|}{\multirow{3}{*}{\begin{tabular}[c]{@{}l@{}}\textbf{Printer 1}:\\ Ultimaker\\ 2+ CONNECT\end{tabular}}} & \multicolumn{1}{c|}{Pre-Print}  & \multicolumn{1}{c|}{00:03:18} & \multicolumn{1}{c|}{12.28} & \multicolumn{1}{c|}{0}    & \multicolumn{1}{c|}{\multirow{3}{*}{0.197}} \\
			
			\multicolumn{1}{|l|}{}                           & \multicolumn{1}{c|}{Print}      & \multicolumn{1}{c|}{00:35:05} & \multicolumn{1}{c|}{77.93} & \multicolumn{1}{c|}{0.17} & \multicolumn{1}{c|}{}                       \\
			
			\multicolumn{1}{|l|}{}                           & \multicolumn{1}{c|}{Post-Print} & \multicolumn{1}{c|}{00:05:52} & \multicolumn{1}{c|}{0.52}  & \multicolumn{1}{c|}{0}    & \multicolumn{1}{c|}{}                       \\ \hline
			\multicolumn{1}{|l|}{\multirow{3}{*}{\begin{tabular}[c]{@{}l@{}}\textbf{Printer 2}:\\ Creality\\ K1 MAX\end{tabular}}} & \multicolumn{1}{c|}{Pre-Print}  & \multicolumn{1}{c|}{00:04:14} & \multicolumn{1}{c|}{14.18} & \multicolumn{1}{c|}{0}    & \multicolumn{1}{c|}{\multirow{3}{*}{0.081}} \\
			\multicolumn{1}{|l|}{}                           & \multicolumn{1}{c|}{Print}      & \multicolumn{1}{c|}{00:09:06} & \multicolumn{1}{c|}{22.06} & \multicolumn{1}{c|}{0.07} & \multicolumn{1}{c|}{}                       \\
			\multicolumn{1}{|l|}{}                           & \multicolumn{1}{c|}{Post-Print} & \multicolumn{1}{c|}{00:00:10} & \multicolumn{1}{c|}{0.1}   & \multicolumn{1}{c|}{0}    & \multicolumn{1}{c|}{}                       \\ \hline
			\multicolumn{1}{|l|}{\multirow{3}{*}{\begin{tabular}[c]{@{}l@{}}\textbf{Printer 3}:\\ Prusa MK4\end{tabular}}} & \multicolumn{1}{c|}{Pre-Print}  & \multicolumn{1}{c|}{00:04:47} & \multicolumn{1}{c|}{15.73} & \multicolumn{1}{c|}{0}    & \multicolumn{1}{c|}{\multirow{3}{*}{0.105}} \\
			\multicolumn{1}{|l|}{}                           & \multicolumn{1}{c|}{Print}      & \multicolumn{1}{c|}{00:10:40} & \multicolumn{1}{c|}{34.29} & \multicolumn{1}{c|}{0.09} & \multicolumn{1}{c|}{}                       \\
			\multicolumn{1}{|l|}{}                           & \multicolumn{1}{c|}{Post-Print} & \multicolumn{1}{c|}{00:00:46} & \multicolumn{1}{c|}{1.02}  & \multicolumn{1}{c|}{0}    & \multicolumn{1}{c|}{}                       \\ \hline
		\end{tabular}%
	}
	\label{tab:tab2}
\end{table*}

Equation (\ref{eq:eq3}) calculates the energy cost based on time and power consumption, converting the 3D printing time (production) into energy costs using the power consumption rate and energy price. The following equation calculates the total production cost of a single ring by adding up all the individual cost components:
\begin{equation}
	C_{production} = C_{material} + C_{time}
\end{equation}

Table \ref{tab:tab2} summarises the total production cost per ring for each printer, including the production time, power consumption and material usage in the pre-print, print and post-print phases. To calculate the operational costs of running a Shopify-based web shop there are several components to considered. Shopify's basic plan, which provides essential e-commerce functionality, costs € 29.00 per month. This includes hosting, SSL certification, and basic e-commerce features. In addition, Shopify charges a transaction fee of 2\% per sale for using external payment providers. When distributing these fixed costs over an average monthly production volume of 100 rings, the \textbf{\textit{Web Shop Operational Costs}} are € 0.29.

The Cloud Crafting Platform, deployed on Microsoft Azure, incurs costs based on resource consumption and service usage. Using the Azure pricing calculator, the monthly cost to run the complete platform architecture includes: \textbf{\textit{API Gateway}} (€ 0.421 per million calls), \textbf{\textit{Load Balancer}} (€ 0.0225 per hour), and the core services running on Azure App Service (€ 0.149 per hour), resulting in approximately € 175 per month for the entire infrastructure. Distributed over an average monthly production volume of 100 rings, the \textbf{\textit{Microsoft Azure Cloud Infrastructure Costs}} are € 1.75 per ring. The following equation can be used to calculate the total production costs per ring including the \textbf{\textit{Web Shop Operational Costs}}, \textbf{\textit{Microsoft Azure Cloud Infrastructure Costs}} and \textbf{\textit{Production Costs}}:
\begin{equation}
	C_{total} = C_{webshop} + C_{cloud} + C_{production}
\end{equation}

\noindent where:
\begin{table}[h]
	\resizebox{\columnwidth}{!}{%
		\begin{tabular}{ll}
			$C_{webshop}$    & is the \textbf{\textit{Web Shop Operational Costs}}\vspace{3pt} \\
			\multirow{2}{*}{$C_{cloud}$}    & is the \textbf{\textit{Microsoft Azure Cloud}} \\
			& \textbf{\textit{Infrastructure Costs}}\vspace{3pt} \\
			$C_{production}$ & is the \textbf{\textit{Production Costs}} per ring\vspace{3pt} \\
		\end{tabular}%
	}
\end{table}

To sum it up, the total cost per ring, combining web shop costs (€ 0.29), cloud costs (€ 1.75), and production costs (depending on the printer) are: 
\begin{itemize}
	\item[]$C_{total, printer1} = 0.29 + 1.75 + 0.197 = $ € $ 2.237$
	\item[]$C_{total, printer2} = 0.29 + 1.75 + 0.081 = $ € $ 2.121$
	\item[]$C_{total, printer3} = 0.29 + 1.75 + 0.105 = $ € $ 2.145$
\end{itemize}

\subsection{Discussion}

The cost-benefit analysis demonstrates the economic viability of the Cloud Crafting Platform and the MaaS approach to additive manufacturing. The total cost per ring, including web shop costs (€ 0.29), cloud platform costs (€ 1.75) and production costs (between € 0.081 and € 0.197), is between € 2.121 and € 2.237, depending on the printer model used. With a reasonable market price of €10-15 per ring, this results in a significant profit margin of around 400-600\%, making it attractive to all stakeholders. 

Scaling up production to 100 rings per month, the Total Cost of Ownership (TCO) would be between € 212.10 and € 223.70. With a potential revenue of € 1,000 (assuming a selling price of € 10 per ring), this generates a monthly profit of approximately € 776.30 to  €787.90. This profit can be shared between the four service-providing stakeholders: the web shop operator, the cloud platform operator, the CAD model designer, and the 3D printer operator. 

Using a weighted distribution model based on infrastructure investment, operational responsibility, and ongoing commitment, each stakeholder could receive the following profit share per month:
\begin{itemize}
	\item\textbf{Cloud Crafting Platform Operator (40\%)}: receives the highest share due to platform maintenance and core service delivery.\\
	\textbf{Total share per month}: € 310.52-315.16
	
	\item\textbf{3D Printer Operator (30\%)}: receives the second highest share for providing equipment and expertise including managing physical production and quality control.\\
	\textbf{Total share per month}: € 232.89-236.37
	
	\item\textbf{Web Shop Operator (20\%)}: receives a share for managing the customer interface, sales, customer support and marketing.\\
	\textbf{Total share per month}: € 155.26-157.58
	
	\item\textbf{CAD Model Designer (10\%)}: receives a share for creating the initial product designs (one-off contribution per product).\\
	\textbf{Total share per month}: € 77.63-78.79
\end{itemize}
\vspace{10pt}

The Creality K1 MAX (\textbf{Printer 2}) demonstrates the most cost-effective production metrics, with the lowest combined material and energy costs (€ 0.081 per ring) and the fastest production time (13:30 minutes), making it particularly suitable for high-volume production scenarios. The MaaS approach not only proves economically viable, but also offers significant resource efficiency and scalability benefits. The ability to produce on-demand eliminates inventory costs and reduces waste, while the distributed manufacturing model enables local production, reducing transportation costs and environmental impact. In addition, the profit-sharing model creates a sustainable ecosystem that incentivises all stakeholders to participate in and contribute to the success of the Cloud Crafting Platform.

\vspace{-10pt}
\section{\uppercase{Conclusions}}
\label{sec:sec5}

This paper presents a comprehensive analysis of a Cloud Crafting Platform that enables MaaS through additive manufacturing technologies. Our research demonstrates both the technical feasibility and economic viability of connecting web shops to local 3D printing facilities through a cloud-based platform. The prototype implementation, with a SOA architecture deployed on Microsoft Azure, successfully integrates all the components required for end-to-end manufacturing services: from online product purchase to local additive production. The testbed setup included three different 3D printer models and provided valuable insights into the operational characteristics and cost structures of on-demand production. 

The cost-benefit analysis shows a compelling business case for all stakeholders. With a total production cost of approximately € 2.24 per ring and a retail price of € 10, the platform generates sufficient margins to sustain a profitable ecosystem. The weighted profit-sharing model (40\% cloud crafting platform operator, 30\% 3D printer operator, 20\% web shop operator, and 10\% CAD designer) ensures fair compensation based on investment levels and operational responsibilities.

In future work we plan to explore the integration of additional manufacturing technologies beyond 3D printing, enhanced quality control mechanisms, and optimisation of the profit-sharing model based on real-world implementation data. The successful demonstration of this platform contributes to the ongoing evolution of distributed manufacturing systems and provides a blueprint for the practical implementation of MaaS.

\section*{\uppercase{Acknowledgements}}

Research leading to these results has received funding from the Digital Innovation Hub Süd (DIH - Süd) for the innovation project RePro3D funded by DIH SÜD GmbH 2023 - 2024.

\vspace{-10pt}
\bibliographystyle{apalike}
{\small
	\bibliography{example}}

\end{document}